\documentclass[prl,reprint,superscriptaddress,nofootinbib,amsmath,subcaption]{revtex4-1}
\usepackage[dvips]{graphics}
\usepackage{subfigure}
\usepackage{epstopdf}
\usepackage{lineno,hyperref}
\usepackage{graphicx}
\usepackage{amssymb}
\usepackage{bm}
\usepackage{color}
\usepackage{multirow}
\usepackage{dcolumn}
\usepackage{makecell}

\newcommand{\dm}[1]{\Delta m^2_{#1}}
\newcommand{\dcp}{\delta_{CP}}

\newcommand{\nova}{NO$\nu$A}

\newcommand{\pme}{P_{\mu e}}

\newcommand{\pmebar}{\overline{P}_{\mu e}}

\newcommand{\adual}{R_{\text{dual}}}

\newcommand{\epsme}{\varepsilon_{\mu e}}

\newcommand{\be}{\begin{equation}}
\newcommand{\ee}{\end{equation}}

\begin{document}

\title{Neutrino oscillations at dual baselines}
\author{Minseok Cho}
\affiliation{Department of Physics, KAIST, Daejeon 34141, Korea}
\author{YeolLin ChoeJo}
\affiliation{Department of Physics, KAIST, Daejeon 34141, Korea}
\author{Hye-Sung Lee}
\affiliation{Department of Physics, KAIST, Daejeon 34141, Korea}
\author{Young-Min Lee}
\affiliation{Department of Physics, KAIST, Daejeon 34141, Korea}
\author{Sushant K. Raut}
\affiliation{Center for Theoretical Physics of the Universe, IBS, Daejeon 34051, Korea}
\date{July 2019}

\begin{abstract}
Beam neutrino oscillation experiments typically employ only one detector at a certain baseline, apart from the near detector that measures the unoscillated neutrino flux at the source.
Lately, there have been discussions of having detectors at two different baselines in one of the future long-baseline neutrino oscillation experiments.
We study the potential advantage of a general dual-baseline system and perform analysis with a specific example of the envisioned T2HKK experiment.
We introduce a new parameter to exploit the correlation between the oscillations at both baselines, and show how it can help in determining the mass hierarchy and the CP phase in the neutrino sector.
Our study and findings can be generically used for any dual-baseline system.
\end{abstract}

\keywords{Neutrino oscillations, dual detector}
\maketitle

\subsubsection{Introduction}
Neutrino oscillations have been crucial in revealing the neutrino sector, which is one of the least explored parts of the Standard Model (SM) extended by neutrino masses.
Although many properties of the neutrino sector have been measured by now, there are still unmeasured ones including the CP phase and the mass hierarchy.
To determine these and precisely measure other parameters in the neutrino oscillation experiments, careful designs and analyses are called for.

Typical neutrino beam experiments explore the probability of neutrino oscillations at a specific baseline ($L$) as a function of the $L/E$ (baseline over energy) along with other neutrino properties relevant to oscillations.
As only a tiny fraction of the neutrinos can interact with the detector, we can use the same beam with another detector at a different baseline after further oscillation.
The proposed T2HKK experiment with a neutrino source at Tokai and two detectors -- one at Kamioka, Japan and the other at Mt. Bisul, Korea is a prime example of a setup with complementary information from the shorter ($295$ km) and longer ($1100$ km) baselines.
Various physics potentials of this experiment have been discussed in the literature \cite{Hagiwara:2005pe,Hagiwara:2006vn,Hagiwara:2011kw,Hagiwara:2016qtb,Fukasawa:2016lew,Abe:2016ero,Liao:2016orc,Ghosh:2017ged,Raut:2017dbh,Choubey:2017cba,Abe:2017jit,Ghosh:2017lim,Chakraborty:2017ccm}.

In this paper, we study a general dual-baseline system, exploiting the correlation between the oscillations at two baselines.
We introduce a new parameter designed to use the correlations of the data at both detectors and investigate advantages of a dual-baseline system in determining the neutrino oscillations parameters and also in finding potential new physics.
As a specific illustrative example, we will consider the case of the T2HKK setup although our study can be applied to broad setups.

\subsubsection{Neutrino mixing}
The PMNS matrix $U$ that relates the experimentally observable flavor eigenstates and propagating mass eigenstates of the neutrino is parametrized by three mixing angles $\theta_{12}$, $\theta_{13}$, $\theta_{23}$ and the CP-violating phase $\dcp$ as  
\begin{widetext}
\begin{equation}
U = \left( \begin{array}{ccc} c_{12} c_{13} & s_{12} c_{13} & s_{13} e^{-i\dcp} \\ -s_{12} c_{23} - c_{12} s_{23} s_{13} e^{i\dcp} & c_{12} c_{23} - s_{12} s_{23} s_{13} e^{i\dcp} & s_{23} c_{13} \\ s_{12} s_{23} - c_{12} c_{23} s_{13} e^{i\dcp} & -c_{12} s_{23} - s_{12} c_{23} s_{13} e^{i\dcp} & c_{23} c_{13} \end{array} \right) ~,
\end{equation}
\end{widetext}
following the PDG convention of parametrization~\cite{Tanabashi:2018oca}. 
The oscillation probabilities also depend on the mass-squared differences $\dm{21}$ and $\dm{31}$. The values of these parameters serve as discriminators between models of new physics that predict the existence of tiny non-zero neutrino masses.
Further, these scenarios can also affect oscillations through non-standard interactions (NSIs).
Consequently there is a worldwide effort by the neutrino physics community towards measuring the oscillation parameters. 

Based on global fits of world neutrino data~\cite{Esteban:2018azc}, typical parameter values close to the global best-fit are $\sin^2\theta_{12} = 0.3$, $\sin^2 2\theta_{13} = 0.085$, $\sin^2\theta_{23} = 0.5$, $\dm{21} = 7.5\times 10^{-5} \textrm{eV}^2$, $|\dm{31}| = 2.6\times 10^{-3} \textrm{eV}^2$. The sign of $\dm{31}$ or neutrino mass hierarchy and the value of $\dcp$ are the main unknown parameters. The measurement of these parameters as well as probes of new physics will continue at the current oscillation experiments such as T2K~\cite{Itow:2001ee}, \nova~\cite{Ayres:2004js}, SK~\cite{Fukuda:1998mi} and IceCube~\cite{Ahrens:2002dv} as well as at the proposed experiments such as T2HK~\cite{Abe:2015zbg}, T2HKK~\cite{Abe:2016ero}, DUNE~\cite{Acciarri:2015uup}, ICAL~\cite{Kumar:2017sdq} and JUNO~\cite{Djurcic:2015vqa}. The synergy between various oscillation channels at different baselines $L$ and energies $E$ at these experiments is crucial in order to break degeneracies in the parameter space.

\subsubsection{Dual ratio}
The $\nu_\mu \to \nu_e$ oscillation probability can be expressed as~\cite{Akhmedov:2004ny}
\begin{eqnarray}
\pme & \approx & F_0 \frac{\sin^2 [(1-phA)\Delta]}{(1-phA)^2} \label{eq:pme} \\
& + & h F_1 \cos[\Delta+ph\dcp] \frac{\sin[A \Delta]}{A} \frac{\sin [(1-phA)\Delta]}{(1-phA)} ~, \nonumber
\end{eqnarray}
where $\Delta=|\dm{31}|L/4E$ and $F_{0,1}$ are constants depending only on the oscillation parameters apart from $\dcp$. We have introduced the binary variables $p=\pm1$ for neutrinos (antineutrinos) and $h=\pm1$ for normal hierarchy, NH (inverted hierarchy, IH).
$A = 2\sqrt{2} G_F n_e E/\dm{31}$  is the magnitude of the dimensionless matter effect term, which depends on the local electron density $n_e$.

We assume the matter effect $A$ is small (which is also valid in the T2HKK setup where $A \sim 0.05$ at the peak energy).
Expanding Eq.~\eqref{eq:pme} up to the first order in $A$, we get
\begin{eqnarray}
\pme & \approx & F_0 \sin^2\Delta + h F_1 \Delta \cos[\Delta+ph\dcp] \sin\Delta \label{eq:pmeapprox} \\
& + & pA T_\Delta (2hF_0 \sin\Delta + F_1 \Delta \cos[\Delta+ph\dcp]) ~, \nonumber
\end{eqnarray}
where $T_\Delta = \sin\Delta-\Delta\cos\Delta$.
The difference between neutrino and antineutrino oscillation probabilities gives a measure of CP violation,
\begin{eqnarray}
 \pme - \pmebar & \approx & -2 F_1 \Delta \sin^2\Delta \sin\dcp  \label{eq:Pdifference} \\
 & + & 2 A F_1 \Delta T_\Delta \cos\Delta \cos\dcp + 4 h A F_0 T_\Delta \sin\Delta ~. \nonumber
\end{eqnarray}
The first term is a measure of intrinsic CP violation induced by $\dcp$ and is independent of matter effects, while the second and third terms are even functions of $\dcp$ and represent matter-induced CP violation.
Ref.~\cite{Bernabeu:2018use} discusses the behavior of these terms under the T and CPT symmetries.
Further, the third term is the only hierarchy dependent term. 

The maxima and minima of $\pme$ and $\pmebar$ depend on $\Delta$ (through its $L/E$ dependence), but also vary slightly with $\dcp$ and the matter term $A$, since these quantities shift the oscillation phase as seen in Eq.~\eqref{eq:pme}.
In Eq.~\eqref{eq:pme}, the $F_0$ term is dominant over the $F_1$ term and, for a sufficiently small $A$, the maxima (minima) are basically determined by $\Delta = n\pi/2$ with $n$ odd (even).
The $\pme - \pmebar$ of Eq.~\eqref{eq:Pdifference} at various $\pme$, $\pmebar$ extrema are
\begin{eqnarray}
 \textrm{1st max } (\Delta=\pi/2) &:& 4hAF_0 - F_1\pi\sin\dcp ~, \nonumber \\
 \textrm{1st min } (\Delta=\pi) &:& - 2AF_1\pi^2\cos\dcp ~, \\
 \textrm{2nd max } (\Delta=3\pi/2) &:& 4hAF_0 - 3F_1\pi\sin\dcp ~, \nonumber \\
 \textrm{2nd min } (\Delta=2\pi) &:& - 8AF_1\pi^2\cos\dcp ~. \nonumber
\end{eqnarray}

\begin{figure}[t]
\centering
\includegraphics[width=0.48\textwidth]{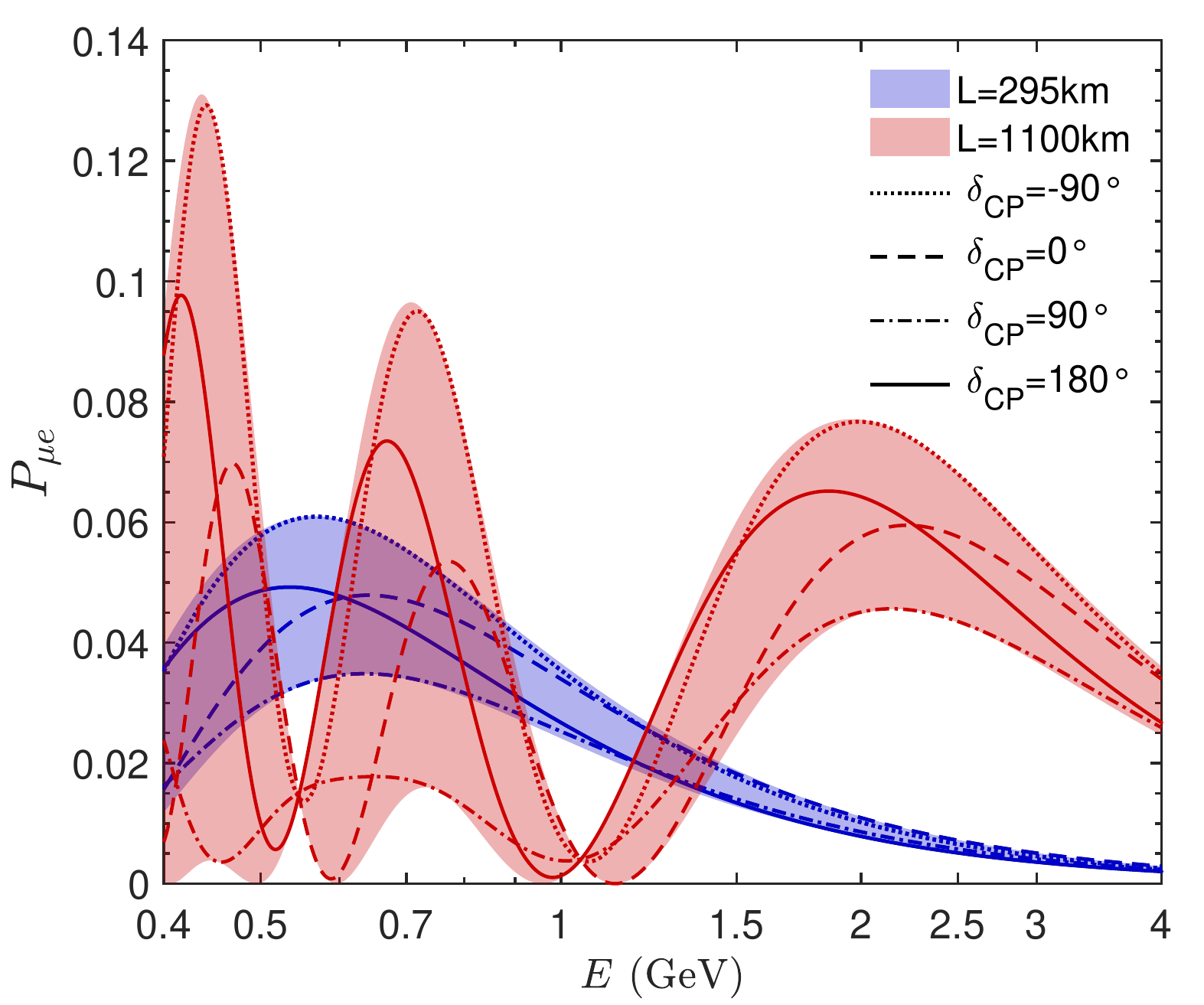}
\caption{$\pme$ at 295 km (blue) and 1100 km (red) for various values of $\dcp$. The shaded regions cover entire $\dcp$ range ($0^\circ - 360^\circ$). The maxima/minima for the $\pmebar$ are located at similar energy values.}
\label{fig:probflux}
\end{figure}

We define a new parameter
\begin{equation}
\adual = \frac{P_{\mu e, 1} - \overline{P}_{\mu e, 1}}{P_{\mu e, 2} - \overline{P}_{\mu e, 2}} ~,
\label{eq:anew}
\end{equation}
and call it `dual ratio', which is the ratio of $\pme - \pmebar$ at a shorter baseline ($L_1$) and at a longer baseline ($L_2$).\footnote{An alternative definition could be $\frac{(P_{\mu e, 1} - \overline{P}_{\mu e, 1}) / (P_{\mu e, 1} + \overline{P}_{\mu e, 1})}{(P_{\mu e, 2} - \overline{P}_{\mu e, 2}) / (P_{\mu e, 2} + \overline{P}_{\mu e, 2})}$, that is the ratio of the typical $A_{CP}$ at the two baseline.}
This measures the correlation between the oscillations at two baselines, and equals $1$ when $L_2 = L_1$.

From Eq.~\eqref{eq:Pdifference}, we see an invariance of
\begin{equation}
h \to -h , ~ \dcp \to 180^\circ+\dcp ~\Longrightarrow ~\adual \to \adual .
\label{eq:degeneracy}
\end{equation}
This is consistent with a well-known degeneracy between \{NH, $\dcp=+90^\circ$\} and \{IH, $\dcp=-90^\circ$\} from experiments such as \nova\ and T2K~\cite{Prakash:2012az}.

$\adual$ can be written conveniently as a simple analytic expression at the energies at which $\pme$, $\pmebar$ extrema occur.
For instance, consider the case the $L_1$ has the first maximum ($\Delta = \pi/2$) at an energy $E$.
The $L_2$ would have the second minimum ($\Delta = 2\pi$) at the same energy if $L_2 \approx 4 L_1$ as $\Delta(\text{at } L_1) / \Delta(\text{at } L_2) = L_1 / L_2 \approx (\pi/2) / (2\pi) = 1/4$, and the $\adual$ is given by
\begin{equation}
\adual \approx -\frac{4hA(L_1) F_0 - F_1\pi\sin\dcp}{8A(L_2) F_1\pi^2\cos\dcp} ~.
\label{eq:anewatpeak}
\end{equation}
In the following, we will see how the $\adual$ can be used to extract important information with an example of the T2HKK setup.

\subsubsection{Dual ratio at T2HKK setup}
Now we focus our analysis on the specific T2HKK setup.
The neutrino flux from J-PARC is designed to peak at the first maximum ($\Delta=\pi/2$) of the T2K or T2HK baseline ($L_1 = 295$ km), which is around $0.6$ GeV.
The second baseline ($L_2 = 1100$ km) was chosen so that the second maximum ($\Delta=3\pi/2$) is covered; the second minimum ($\Delta=2\pi$) occurs there too at $E=0.57$ GeV.
(See Fig.~\ref{fig:probflux}.)
Studies show that adding the second baseline is more useful in determining $\dcp$ than the T2HK setup using only one baseline \cite{Abe:2016ero}.

For $E = 0.57$ GeV in the T2HKK setup, Eq.~\eqref{eq:anewatpeak} gives
\begin{equation}
\adual \approx -0.05 h \sec\dcp + 0.14 \tan\dcp ~.
\label{eq:T2HKK}
\end{equation}
All plots in this article use the numerically generated exact values \cite{Huber:2004ka,Huber:2007ji}.

Some interesting features of the new parameter $\adual$ for $E = 0.57$ GeV in Eq.~\eqref{eq:T2HKK} are as follows: 
\begin{enumerate}
\item[(i)] The first (second) term is proportional to $h \sec\dcp$ ($\tan\dcp$), which is $\dcp$-even ($\dcp$-odd), matter-independent (matter-dependent), and hierarchy-dependent (hierarchy-independent).
Both $\sec\dcp$ and $\tan\dcp$ diverge at $\dcp = \pm90^\circ$, which is interesting as we expect a very good resolution in $\dcp$ from the $\adual$ near $\dcp=-90^\circ$ which is favored by the current data \cite{NOvA:2018gge,Abe:2018wpn}.
\item[(ii)] When $h \to -h$ and $\dcp \to -\dcp$, we get $\adual \to -\adual$.
This fact can be used to break the mass hierarchy--$\dcp$ degeneracy.
We can use the sign of $\adual$ to resolve this degeneracy when $\dcp$ is close to $-90^\circ$.
\end{enumerate}

\begin{figure}[t]
{
\centering
\includegraphics[trim=0 0 450 0,clip,width=0.48\textwidth]{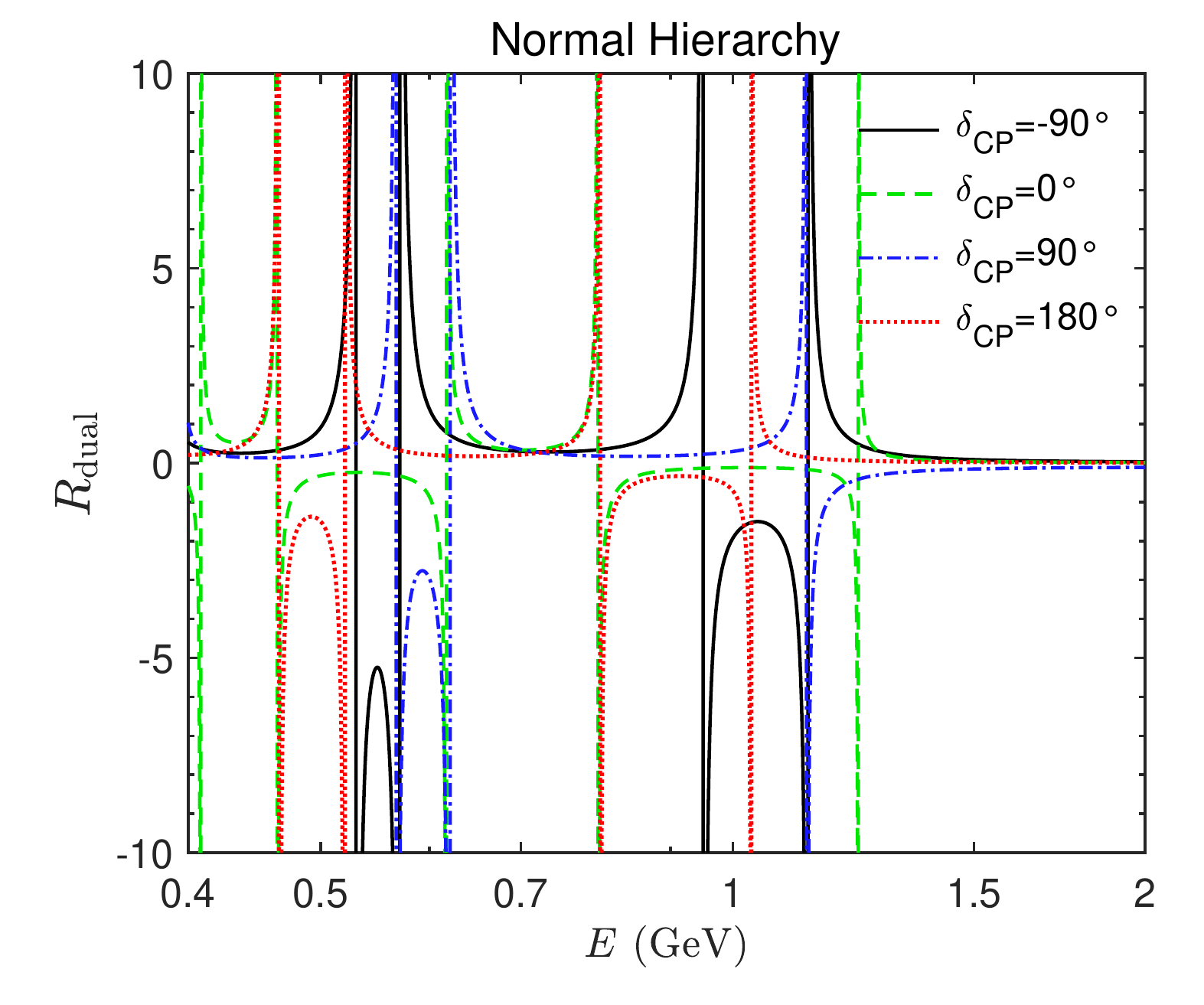}
\caption{$\adual$ vs $E$ at T2HKK for several values of $\dcp$, assuming NH. Each $\dcp$ shows a characteristic series of spikes. The spikes would allow a high resolution in identifying the energy values.}
\label{fig:asymm}
}
\end{figure}

\begin{figure*}[t]
{
\centering
\includegraphics[width=0.96\textwidth]{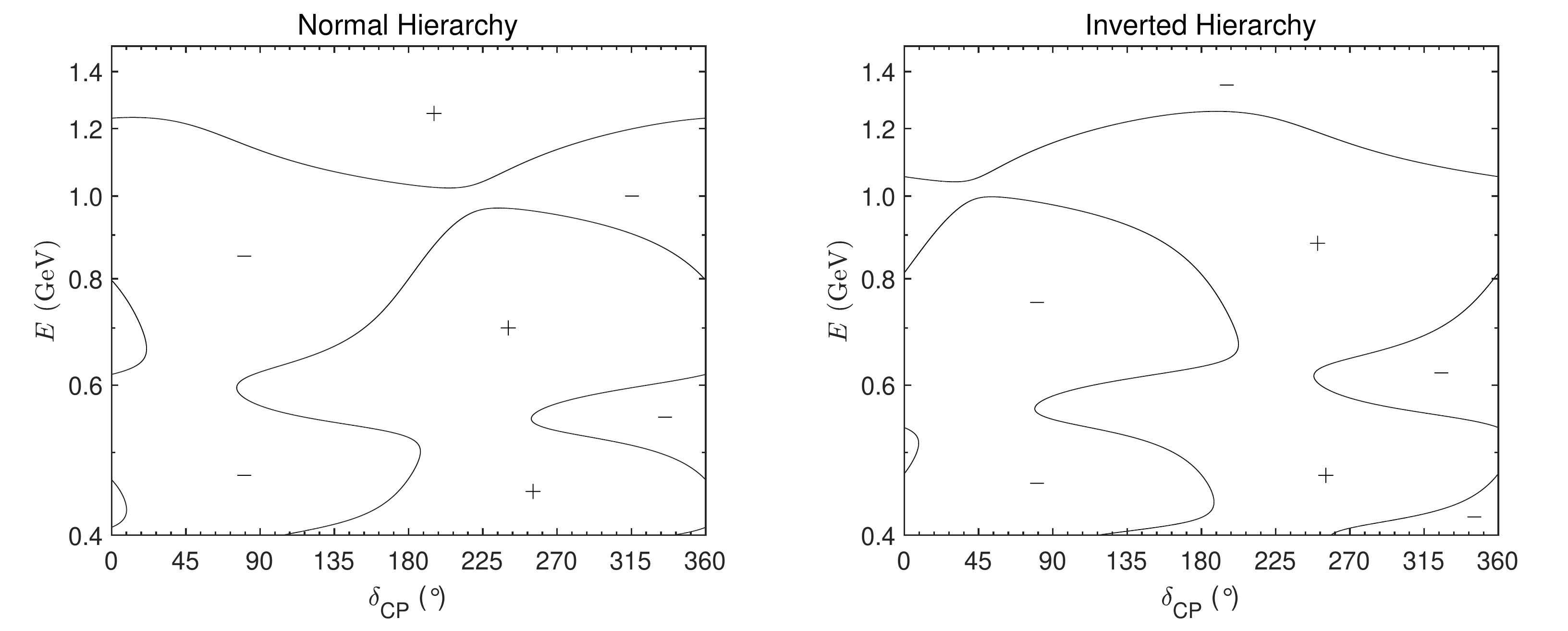}
\caption{
Contours indicate the points where the spikes of Fig.~\ref{fig:asymm} at T2HKK occur.
At these points, $\adual$ diverges.
The set of energies at which the diverging $\adual$ are observed thus can be used to read-off the value of $\dcp$ using these plots as a reference chart.
The mass hierarchy--$\dcp$ degeneracy \eqref{eq:degeneracy} is clear between the NH and IH, but it can be resolved by comparing the sign of the denominator of the $\adual$, which is shown in each closed region.
}
\label{fig:kzeros}
}
\end{figure*}

\begin{figure}[b]
{
\centering
\includegraphics[width=0.48\textwidth]{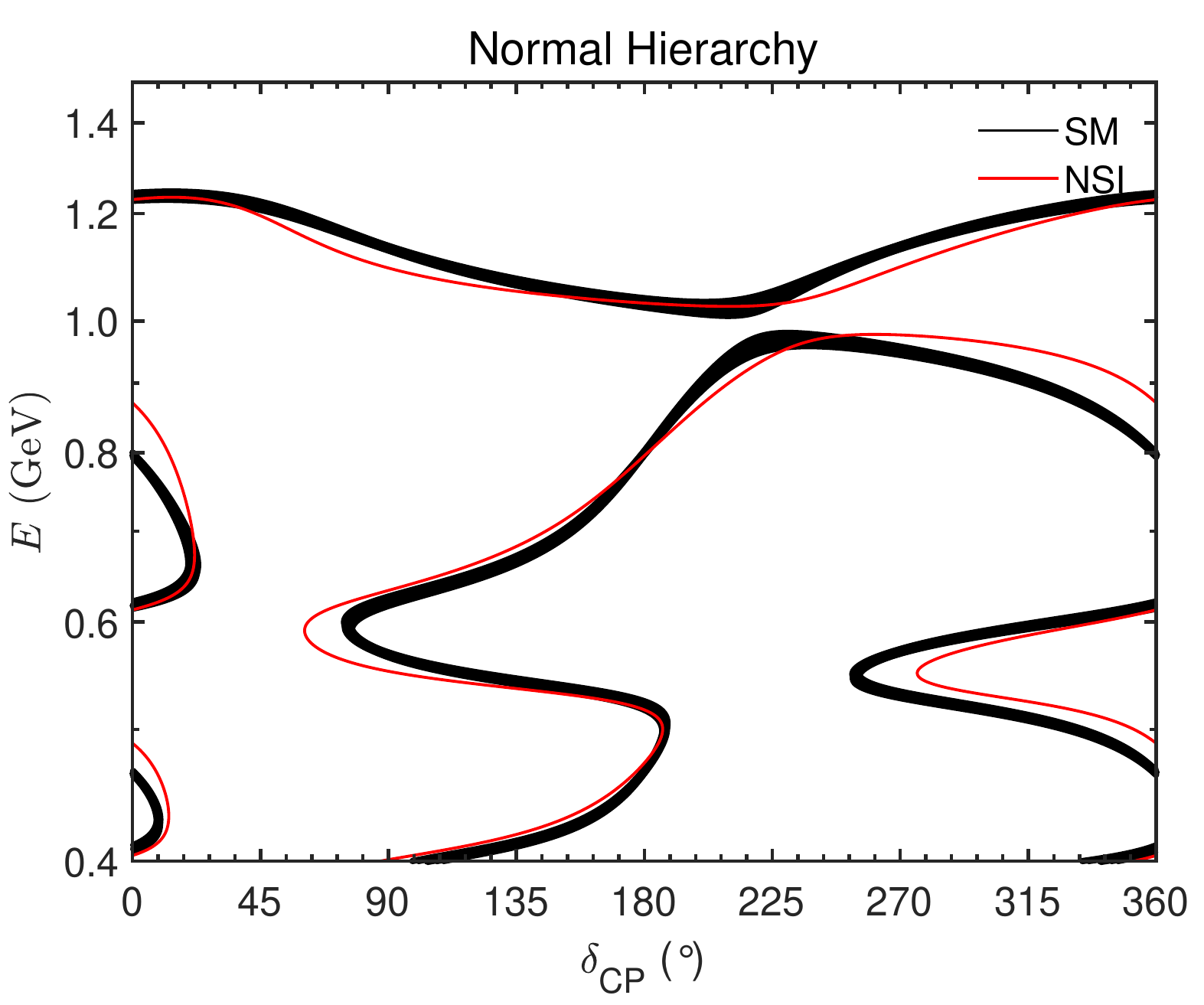}
\caption{Locations of the $\adual$ spikes, assuming the NSI $\epsme=0.07$ (thin red contours). The standard oscillation is the spread over the oscillation parameters in their projected $1\sigma$ ranges (black band).
Spikes found outside of the black band would suggest a new physics.}
\label{fig:nsizeros}
}
\end{figure}

We now discuss the $\adual$ over a wide energy range.
Figure~\ref{fig:asymm} shows the energy dependence of $\adual$ for several values of $\dcp$.
Each value of $\dcp$ gives a characteristic series of spikes at the energies where $\adual$ diverge.
Since the first zero for the 1100 km baseline lies around 1.2 GeV, there are no spikes above this energy.
The diverging spikes would allow a high resolution in identifying the energies.

Figure~\ref{fig:kzeros} shows the contours where $\adual$ diverges (or its denominator $P_{\mu e, 2} - \overline{P}_{\mu e, 2}$ is zero) in the $E$--$\dcp$ plane.
The $+$, $-$ signs in the plane show the sign of its denominator in the region.
We can use Fig.~\ref{fig:kzeros} as the `standard chart' to compare the diverging positions to the data and deduce the $\dcp$ and mass hierarchy.
The high resolution of the energies with the spikes allows an accurate decision.
Even if some points are somewhat affected by the mass hierarchy--$\dcp$ degeneracy \eqref{eq:degeneracy}, it can be resolved by taking an approach described in feature (ii).
However, it would be simpler to determine the hierarchy by reading the sign of the denominator and compare to the signs in the plots.

We can generalize the discussion here to any dual-baseline setup, where $\adual$ diverges in some energies.
The resulting spectrum of spikes in $\adual$ will be characteristic of a specific value of $\dcp$ and hierarchy.

\subsubsection{Non-standard interactions}
A similar analysis can be performed for the oscillation probabilities in the presence of non-standard interactions (NSIs).
As an illustrative example we consider the case of one non-zero NSI parameter $\epsme$.
(For the exact definition of $\epsme$ and the current constraints on it, see Refs.~\cite{Ohlsson:2012kf,Farzan:2017xzy}.)
We assume the value $\epsme=0.07$, which is the expected sensitivity reach of DUNE \cite{Coloma:2015kiu}.
(T2HK using one baseline is expected to have lower sensitivity, only up to around $\epsme=0.20$.)
Figure~\ref{fig:nsizeros} shows where the spikes would appear for the given NSI as well as for standard oscillations.
In the standard case we have also varied the oscillation parameters in their $1\sigma$ allowed ranges based on projected sensitivities from the disappearance data~\cite{Abe:2016ero}, resulting in a thick band.
If the spikes are found outside of this band, they would suggest a new physics.
The proposed HK detectors will have an impressive energy resolution of around 20 MeV \cite{Abe:2015zbg}, and it would be possible to tell the difference fairly accurately.

\subsubsection{Summary and outlooks}
In this article, we investigated the potential advantages of a dual-baseline neutrino oscillation system.
Using the correlation of the oscillations at two baselines would be a key to take full advantage of a dual-baseline system, yet there are few (if not none) studies in the literature trying this so far.
We introduced a simple parameter, $\adual$, that can measure the correlation of the two and used it to discuss the neutrino oscillations at the dual-baseline system.

We studied, for a specific example, the T2HKK experiment of two baselines (295 km and 1100 km) in our analysis.
Our study is timely as there are ongoing active discussions about the possibility and capability of the T2HKK.
We found that our approach using the correlation allows one to accurately measure the series of energies which is crucial in deciding the value of the CP violating phase $\dcp$.
(Especially, it promises a remarkably high precision for the hinted $\dcp \sim -90^\circ$ near the peak energy of the first baseline.)
We also found the potential mass hierarchy--$\dcp$ degeneracy can be resolved.

We presented a reference chart one can use to read-off the $\dcp$ value directly once the experimental data is given.
The reference chart can be also used to reveal if there is a new physics affecting the neutrino oscillations.
We illustrated this with one specific example, which shows an intriguing result.
Typical $L/E$ dependence may not be valid in the presence of the new physics \cite{Joshipura:2003jh}, and the dual-baseline system with the correlations may play a very useful role.
Dedicated studies for general new physics scenarios are called for, using the correlations of the dual-baseline systems.

There are many potential utilities of our method in broader directions.
The dual-baseline system exploiting the correlations of the oscillations may apply to short-baseline beam experiments and possibly other types of oscillations too.

\vspace{2mm}
{\em Acknowledgements:} This work was supported in part by IBS (Project Code IBS-R018-D1) and NRF Strategic Research Program (NRF-2017R1E1A1A01072736).



\begin{thebibliography}{99}

\bibitem{Hagiwara:2005pe} 
  K.~Hagiwara, N.~Okamura and K.~i.~Senda,
  Phys.\ Lett.\ B {\bf 637}, 266 (2006)
  Erratum: [Phys.\ Lett.\ B {\bf 641}, 491 (2006)]
  doi:10.1016/j.physletb.2006.09.003, 10.1016/j.physletb.2006.04.041
  [hep-ph/0504061].

\bibitem{Hagiwara:2006vn} 
  K.~Hagiwara, N.~Okamura and K.~i.~Senda,
  Phys.\ Rev.\ D {\bf 76}, 093002 (2007)
  doi:10.1103/PhysRevD.76.093002
  [hep-ph/0607255].

\bibitem{Hagiwara:2011kw} 
  K.~Hagiwara, N.~Okamura and K.~i.~Senda,
  JHEP {\bf 1109}, 082 (2011)
  doi:10.1007/JHEP09(2011)082
  [arXiv:1107.5857 [hep-ph]].

\bibitem{Hagiwara:2016qtb} 
  K.~Hagiwara, P.~Ko, N.~Okamura and Y.~Takaesu,
  Eur.\ Phys.\ J.\ C {\bf 77}, no. 3, 138 (2017)
  doi:10.1140/epjc/s10052-017-4684-1
  [arXiv:1605.02368 [hep-ph]].

\bibitem{Abe:2016ero} 
  K.~Abe {\it et al.} [Hyper-Kamiokande Collaboration],
  PTEP {\bf 2018}, no. 6, 063C01 (2018)
  doi:10.1093/ptep/pty044
  [arXiv:1611.06118 [hep-ex]].

\bibitem{Fukasawa:2016lew} 
  S.~Fukasawa, M.~Ghosh and O.~Yasuda,
  Phys.\ Rev.\ D {\bf 95}, no. 5, 055005 (2017)
  doi:10.1103/PhysRevD.95.055005
  [arXiv:1611.06141 [hep-ph]].

\bibitem{Liao:2016orc} 
  J.~Liao, D.~Marfatia and K.~Whisnant,
  JHEP {\bf 1701}, 071 (2017)
  doi:10.1007/JHEP01(2017)071
  [arXiv:1612.01443 [hep-ph]].

\bibitem{Ghosh:2017ged} 
  M.~Ghosh and O.~Yasuda,
  Phys.\ Rev.\ D {\bf 96}, no. 1, 013001 (2017)
  doi:10.1103/PhysRevD.96.013001
  [arXiv:1702.06482 [hep-ph]].

\bibitem{Raut:2017dbh} 
  S.~K.~Raut,
  Phys.\ Rev.\ D {\bf 96}, no. 7, 075029 (2017)
  doi:10.1103/PhysRevD.96.075029
  [arXiv:1703.07136 [hep-ph]].

\bibitem{Choubey:2017cba} 
  S.~Choubey, D.~Dutta and D.~Pramanik,
  Phys.\ Rev.\ D {\bf 96}, no. 5, 056026 (2017)
  doi:10.1103/PhysRevD.96.056026
  [arXiv:1704.07269 [hep-ph]].

\bibitem{Abe:2017jit} 
  Y.~Abe, Y.~Asano, N.~Haba and T.~Yamada,
  Eur.\ Phys.\ J.\ C {\bf 77}, no. 12, 851 (2017)
  doi:10.1140/epjc/s10052-017-5294-7
  [arXiv:1705.03818 [hep-ph]].

\bibitem{Ghosh:2017lim} 
  M.~Ghosh and O.~Yasuda,
  arXiv:1709.08264 [hep-ph].

\bibitem{Chakraborty:2017ccm} 
  K.~Chakraborty, K.~N.~Deepthi and S.~Goswami,
  Nucl.\ Phys.\ B {\bf 937}, 303 (2018)
  doi:10.1016/j.nuclphysb.2018.10.013
  [arXiv:1711.11107 [hep-ph]].

\bibitem{Tanabashi:2018oca} 
  M.~Tanabashi {\it et al.} [Particle Data Group],
  Phys.\ Rev.\ D {\bf 98}, no. 3, 030001 (2018).
  doi:10.1103/PhysRevD.98.030001

\bibitem{Esteban:2018azc} 
  I.~Esteban, M.~C.~Gonzalez-Garcia, A.~Hernandez-Cabezudo, M.~Maltoni and T.~Schwetz,
  JHEP {\bf 1901}, 106 (2019)
  doi:10.1007/JHEP01(2019)106
  [arXiv:1811.05487 [hep-ph]].

\bibitem{Itow:2001ee} 
  Y.~Itow {\it et al.} [T2K Collaboration],
  hep-ex/0106019.

\bibitem{Ayres:2004js} 
  D.~S.~Ayres {\it et al.} [NOvA Collaboration],
  hep-ex/0503053.

\bibitem{Fukuda:1998mi} 
  Y.~Fukuda {\it et al.} [Super-Kamiokande Collaboration],
  Phys.\ Rev.\ Lett.\  {\bf 81}, 1562 (1998)
  doi:10.1103/PhysRevLett.81.1562
  [hep-ex/9807003].

\bibitem{Ahrens:2002dv} 
  J.~Ahrens {\it et al.} [IceCube Collaboration],
  Nucl.\ Phys.\ Proc.\ Suppl.\  {\bf 118}, 388 (2003)
  doi:10.1016/S0920-5632(03)01337-9
  [astro-ph/0209556].

\bibitem{Abe:2015zbg} 
  K.~Abe {\it et al.} [Hyper-Kamiokande Proto- Collaboration],
  PTEP {\bf 2015}, 053C02 (2015)
  doi:10.1093/ptep/ptv061
  [arXiv:1502.05199 [hep-ex]].

\bibitem{Acciarri:2015uup} 
  R.~Acciarri {\it et al.} [DUNE Collaboration],
  arXiv:1512.06148 [physics.ins-det].

\bibitem{Kumar:2017sdq} 
  S.~Ahmed {\it et al.} [ICAL Collaboration],
  Pramana {\bf 88}, no. 5, 79 (2017)
  doi:10.1007/s12043-017-1373-4
  [arXiv:1505.07380 [physics.ins-det]].

\bibitem{Djurcic:2015vqa} 
  Z.~Djurcic {\it et al.} [JUNO Collaboration],
  arXiv:1508.07166 [physics.ins-det].

\bibitem{Akhmedov:2004ny} 
  E.~K.~Akhmedov, R.~Johansson, M.~Lindner, T.~Ohlsson and T.~Schwetz,
  JHEP {\bf 0404}, 078 (2004)
  doi:10.1088/1126-6708/2004/04/078
  [hep-ph/0402175].
  
\bibitem{Bernabeu:2018use} 
  J.~Bernabéu and A.~Segarra,
  JHEP {\bf 1811}, 063 (2018)
  doi:10.1007/JHEP11(2018)063
  [arXiv:1807.11879 [hep-ph]].
  
\bibitem{Prakash:2012az} 
  S.~Prakash, S.~K.~Raut and S.~U.~Sankar,
  Phys.\ Rev.\ D {\bf 86}, 033012 (2012)
  doi:10.1103/PhysRevD.86.033012
  [arXiv:1201.6485 [hep-ph]].

\bibitem{Huber:2004ka} 
  P.~Huber, M.~Lindner and W.~Winter,
  Comput.\ Phys.\ Commun.\  {\bf 167}, 195 (2005)
  doi:10.1016/j.cpc.2005.01.003
  [hep-ph/0407333].

\bibitem{Huber:2007ji} 
  P.~Huber, J.~Kopp, M.~Lindner, M.~Rolinec and W.~Winter,
  Comput.\ Phys.\ Commun.\  {\bf 177}, 432 (2007)
  doi:10.1016/j.cpc.2007.05.004
  [hep-ph/0701187].

\bibitem{NOvA:2018gge} 
  M.~A.~Acero {\it et al.} [NOvA Collaboration],
  Phys.\ Rev.\ D {\bf 98}, 032012 (2018)
  doi:10.1103/PhysRevD.98.032012
  [arXiv:1806.00096 [hep-ex]].

\bibitem{Abe:2018wpn} 
  K.~Abe {\it et al.} [T2K Collaboration],
  Phys.\ Rev.\ Lett.\  {\bf 121}, no. 17, 171802 (2018)
  doi:10.1103/PhysRevLett.121.171802
  [arXiv:1807.07891 [hep-ex]].
  
\bibitem{Ohlsson:2012kf} 
  T.~Ohlsson,
  Rept.\ Prog.\ Phys.\  {\bf 76}, 044201 (2013)
  doi:10.1088/0034-4885/76/4/044201
  [arXiv:1209.2710 [hep-ph]].

\bibitem{Farzan:2017xzy} 
  Y.~Farzan and M.~Tortola,
  Front.\ in Phys.\  {\bf 6}, 10 (2018)
  doi:10.3389/fphy.2018.00010
  [arXiv:1710.09360 [hep-ph]].

\bibitem{Coloma:2015kiu} 
  P.~Coloma,
  JHEP {\bf 1603}, 016 (2016)
  doi:10.1007/JHEP03(2016)016
  [arXiv:1511.06357 [hep-ph]].
  
\bibitem{Joshipura:2003jh} 
  A.~S.~Joshipura and S.~Mohanty,
  Phys.\ Lett.\ B {\bf 584}, 103 (2004)
  doi:10.1016/j.physletb.2004.01.057
  [hep-ph/0310210].

\end{thebibliography}
\end{document}